# Interpretability of machine-learning models in physical sciences

Luca M. Ghiringhelli,
The NOMAD Laboratory at the FHI-MPG and HU, Berlin, Germany

**Status**

Training a supervised machine-learning (ML) model that yields satisfactory predictions (i.e., maps input features into values of the target property, with errors below a threshold perceived as tolerable) on test data that are driven from the same distribution as the training data, is a task that is nowadays almost routinely accomplished. However, the crucial interest is in that the trained model can *generalize*, i.e., it can yield trustable predictions also for test data that are significantly different from the training data. As human beings, i.e., users who are asked to judge and/or trust predictions of a ML model, we need to *understand what the model has learned*. Such innate need is related to the notion of *interpretability* of the ML model. The literature on interpretability is vast[1-7], but the field is pre-paradigmatic, i.e., it has not reached a consensus on what are the fundamental questions and what are the quantities to be measured. Two somewhat contrasting aspects are typically associated to interpretability[1-3]: *transparency* of the model and its (*post hoc) explainability*. Transparency connects to scientific practice, where a phenomenon is felt as understood when a predictive mathematical law is formulated, which is expected to work with no exception, at least in a well-defined *domain of applicability*. Such law is expected to be *simple*, so that our brains can process most if not all of its consequences. *Explainability* refers to the possibility to inspect a perceived "black-box", i.e., a model that is in general too complex to be grasped by the human mind, but that can be investigated, in order to reveal, for instance, which part of the input mostly affected the output. Incidentally, understanding a decision made by a human refers to the post hoc explainability of what happens in our brains, whose detailed mechanics are beyond current grasp, while we can provide reasons on how a decision was reached, typically based on "similar cases"[2]. Understanding interpretability and in particular devising one or a set of consensus *metrics* for assessing the generalizability and trustability of ML model is one crucial next step, or the field might face another "winter" due to a consequent lack of trust in ML applicability.

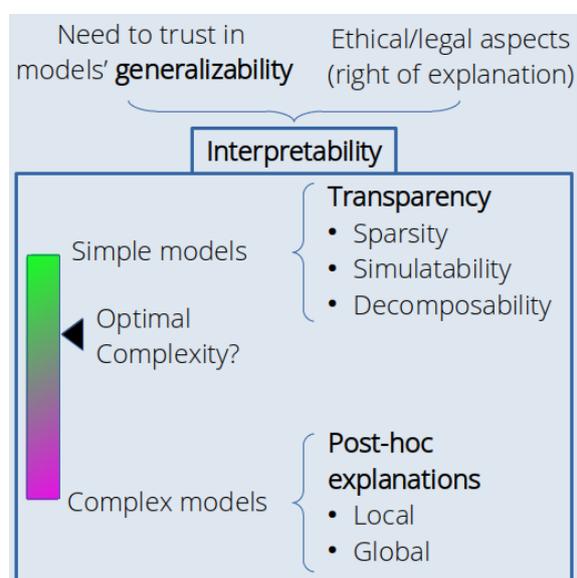

*Summarized view of interpretability in machine learning, elucidating what determines the users' need for interpretability and how the meaning of the word adapts to the complexity of the learned model. Not mentioned in the text, the ethical/legal aspects are felt important when ML-based decisions impact individuals or communities.*

**Current and Future Challenges**

The tools for addressing the interpretability of ML models vary with the complexity of the models[1-6] (see Fig. 1). For simpler models, *transparency* is evaluated, i.e., the ability to read and inspect the model. *Sparse* models[8] and in particular *symbolic inference*[9] naturally provide transparent models as they appear as equations (or inequalities) in terms of functions of input features, which are selected out of a possibly large number of candidates. The interpretation is therefore provided by the identification of which input features govern the modeled phenomenon. Here, the notions of *simulatability* and *decomposability* have been introduced. These are the ability to follow step-by-step how the ML model produces an output from the input and the ability to assign a meaning to each part of a model (e.g., the sign and magnitude of regression coefficients), respectively. An outstanding challenge is to define a rigorous *metric of transparency*, so that models can be objectively compared, similarly and complementarily to the routinely performed, but insufficiently informative, comparison in terms of predictive accuracy.

For more complex model, where transparency is lost, a plethora of *post hoc explanation* tools have been developed[4-7], which are commonly divided into *local* (explanation on how a given single output is obtained) or *global* (typically, visual analysis of how the dataset is represented internally by the model). The focus is in general on a statistical analysis on how input features affect the results. The challenge is here to properly account for the (typically nonlinear) relationship among the input features.

It is highly unsatisfactory that two different interpretability concepts exist depending on the complexity of the trained model. In facts, there is a continuum of complexity between sparse, symbolic models and complex ones (e.g., deep neural networks); the challenge is to seamlessly adapt the complexity of the learned model, and the related interpretability tools, to the intrinsic complexity of the underlying input-features — target-property relationship.

Finally, the importance of *outliers*, datapoints not conforming to the model being learned, needs to be understood. In physical sciences, a wrongly predicted datapoint may be a signal that a different mechanism from the so-far identified features—property relationship is at work.

**Advances in Science and Technology to Meet Challenges**

ML is urgently requested to undergo a paradigm change. Together with prediction accuracy, strategies for assessing the correct model complexity and interpretability metrics, need to be developed. If a simple, symbolic law is the underlying model, a correct ML strategy must be able to recover such exact model. When a more complex, less transparent model is necessary, then the interpretability metric needs to seamlessly adapt to the increased complexity. It should become therefore common practice to compare models in terms not only of their predictive accuracy, but also of their interpretability metric. When applied to the development of scientific (e.g., physical) laws, the purpose of this formidable task is to provide reasons to accept an ML-learned features—property relationship in terms of its consistency with the existing bulk of knowledge, so that the ML model is not felt as a surrogate, until "something better" is found, but as a new scientific law.

In this respect, it is crucial to be able to treat the nonconforming datapoints. Most current ML approaches are built to neglect such datapoints, a.k.a. outliers, while in physical sciences even one single datapoint not complying with the general law is treated with uttermost care, as it could be the harbinger of "new physics". It is therefore desirable that, together with the complexity-aware strategy sketched above, a nonconforming-datapoints strategy is developed. For instance, one may wish to detect different *domains of applicability* of more complex, general models, vs specialized but simpler models. A useful analogy could be thinking at general relativity, which is more general and more complex than classical gravitation. The latter is however very accurate in a well-specified and understood *domain of applicability*. In turn, general relativity is expected to be a special, somewhat simpler, case of a (yet to be developed) quantum-gravity theory. Similarly, in ML the level of complexity of the learned models might need to be adapted to well-defined domains of applicability[10], preferably defined by ML algorithms in a data-driven fashion.

**Concluding Remarks**

In conclusion, ML might have reached its maturity in terms of predictive ability, on data that are statistically similar to the training data. However, it is still in its infancy when it comes i) to generalizability to data significantly different from training data, ii) treatment of "outliers", i.e., data data do not conform to te model being trained, iii) having a unified concept of interpretability that seamlessly applies from the obvious transparency of sparse, symbolic models, to the explainability of complex deep neural networks, and iv) adapting the trained model complexity to the intrinsic complexity of the underlying input feature—property relationship. Hopefully, framing the objective in clear terms will stimulate a focused development of ML techniques, which could promote ML tools to become valuable companions of a scientist, in order to foster future scientific discoveries.

**Acknowledgements**

I acknowledge Jilles Vreeken, Angelo Ziletti, and Matthias Scheffler for insightful discussions. This work received funding from the European Union's Horizon 2020 Research and Innovation Programme (grant agreement No. 676580), the NOMAD laboratory CoE, and ERC:TEC1P (No. 740233).